\documentclass[traditabstract]{aa} % for the abstract without structuration 
\usepackage{graphicx}
\usepackage{natbib}
\usepackage{txfonts}
\def\rs{\rm s}
\def\rs1{\rm s^{-1}}

\def\rcm{\rm cm}
\def\rcm2{\rm cm^{-2}}

\def\c2r{\chi^2_\nu}
\def\chisq{\chi^2}

\def\epo{$E_{\rm p}$ }
\def\epi{\ensuremath{E_{\rm p,i}}}
\def\eiso{\ensuremath{E_{\rm iso}}}
\def\ega{\ensuremath{E_{\rm jet}}}
\def\eiso{\ensuremath{E_{\rm iso}}}
\def\liso{\ensuremath{L_{\rm iso}}}
\def\lpiso{\ensuremath{L_{\rm p,iso}}}
\def\ega{\ensuremath{E_{\gamma}}}
\def\epeiso{$E_{\rm p,i}$ -- $E_{\rm iso}$}
\def\nufnu{$\nu F_{\nu}$ }
\def\epega{$E_{\rm p,i}$ -- $E_{\gamma}$ }
\def\epeisotb{$E_{\rm p,i}$ -- $E_{\rm iso}$ -- $t_{\rm b}$ }
\def\sext{$\sigma$$_{\rm ext}$}
\def\sax{{\it Beppo}SAX }
\def\swift{{\it Swift}}
\def\fermi{{\it Fermi}}
\def\konus{{\it Konus}}
\def\suz{{\it Suzaku}}

\def\apjs{{\it Astrophys. J. Suppl. Ser. }}

\def\omegam{$\Omega$$_{\rm M}$}
\def\omegal{$\Omega$$_{\Lambda}$}
\def\h0{H$_{\rm 0}$~}
\def\tb{t$_{\rm b}$~}
\begin{document}
%
%   \title{Spectrum--energy correlations in Gamma--Ray Bursts confront
%   extremely energetic
   \title{Extremely energetic \fermi{} Gamma--Ray Bursts obey
          spectral energy correlations}
%  \subtitle{I. Overviewing the $\kappa$-mechanism}
   \author{L. Amati
          \inst{1}
          \and
          F. Frontera\inst{1,2}
          \and
          C. Guidorzi\inst{2}
          }
   \institute{INAF - IASF Bologna, via P. Gobetti 101, 40129 Bologna (Italy)
              \email{amati@iasfbo.inaf.it}
         \and
             University of Ferrara, Department of Physics, via Saragat 1, 44100
	     Ferrara (FE)
	     \\
             }
   \date{Submitted June 29, 2009}
% \abstract{}{}{}{}{} 
% 5 {} token are mandatory
  \abstract{
  The origin, reliability and dispersion of the \epeiso{} and other
  spectral energy correlations is a highly debated topic in GRB astrophysics.
      GRB\,080916C, with its huge radiated energy (\eiso $\sim$ 10$^{55}$ erg 
      in the 1 keV -- 10 GeV cosmological rest--frame energy
      band) and its intense GeV emission measured by \fermi,
      gives us a unique opportunity to further investigate this issue.
      We also include in our analysis another extremely energetic event, GRB\,090323,
      more recently detected and localized by \fermi/LAT and showing a radiated
      energy comparable to that of GRB\,080916C in the 1 keV -- 10 MeV energy range.
Based on \konus/WIND and \fermi{} spectral
  measurements,
  we find that both events are fully consistent with the \epeiso{} correlation
  (updated to 95 GRBs with the data available as of April 2009),
  thus further confirming and extending it 
  and pointing against
  a possible flattening or increased dispersion
    at very high energies.
  This also
  suggests that
  the physics behind the emission of peculiarly bright and hard 
  GRBs is the same
  as for medium--bright and soft--weak long events (XRFs), which all follow the
  correlation.
In addition, we find that the normalization of the correlation obtained by considering
these two GRBs and the other long ones for which \epi{} was measured with 
high accuracy by the \fermi/GBM 
are fully
consistent with those obtained by other instruments
(e.g., \sax, \swift, 
\konus/WIND), thus indicating that the correlation is not affected significantly by  
"data truncation" due to detector
thresholds and limited energy bands.
The very recent \fermi/GBM accurate estimate of the peak energy of a
very bright and hard short GRB with measured redshift, GRB\,090510,
provides further and robust evidence that short GRBs do not follow the 
\epeiso{} correlation 
and that the \epeiso{} 
plane can be used to 
discriminate and understand the two classes of events.
Prompted by the extension of the spectrum of GRB\,080916C up to several 
GeVs (in the cosmological rest--frame) without any excess or cut--off, we
also investigated if the evaluation of \eiso{} in the commonly adopted 1 keV -- 10 MeV
energy band may bias the \epeiso{} correlation and/or contribute to its scatter.
By computing 
\eiso{} from 1 keV to 10 GeV, the slope of the correlation becomes 
slightly flatter, while its 
dispersion does not change significantly. 
Finally,
we find that
GRB\,080916C is also consistent with most of 
  the other spectral energy correlations
  derived from it,
with the 
possible exception of the \epeisotb{} correlation.
   }
   {}
   \maketitle
%
%________________________________________________________________

\section{Introduction}

Despite the huge observational and theoretical advances in the
last few years, our understanding of the GRB phenomenon is still affected by relevant 
open issues. Among these, the correlation
between the photon energy at which the \nufnu spectrum (in the cosmological 
rest--frame) of the prompt emission peaks, \epi, and the total radiated energy 
computed by assuming isotropic emission, \eiso, in long GRBs is one of the most 
debated and intriguing. 
Discovered in 2002 based on a sample of BeppoSAX GRBs with 
known redshift \citep{Amati02}, the \epeiso{} correlation was then confirmed and 
shown to hold for all GRBs, soft or bright, with known $z$ and constrained values of \epo and fluence,
with the only exception of the peculiar sub--energetic GRB\,980425 
\citep{Amati06,Amati07}. 
The existence of such a correlation was also supposed 
by Lloyd et al. (2000)\nocite{Lloyd00}
based on the analysis of a sample 
of bright BATSE GRBs
without measured redshift. 
The implications of this observational evidence
can include the physics and geometry of the prompt emission, the identification and 
understanding of different sub--classes of GRBs (e.g., short, sub--energetic), the 
use of GRBs for the estimate of cosmological parameters 
\citep{Amati06,Ghirlanda06,Amati07,Amati08}. 

Thus, testing the \epeiso{} correlation and the other
"spectral energy" correlations derived from it, understanding their origin
and investigating their dispersion and the existence of possible outliers
is a relevant issue for GRB physics and cosmology.
This can be done in three ways: a) by adding new data of GRBs with known redshift 
detected by different instruments, each 
one having its own sensitivity and spectral response and thus covering different 
regions of the \epo{} -- fluence plane \citep{Amati06,Amati06b,Ghirlanda08}; 
b) by verifying its validity with large samples of GRBs with no measured 
redshift \citep{Ghirlanda05a,Ghirlanda08}; c) by studying the behaviour in the 
\epeiso{} plane of peculiar GRBs \citep{Amati07}.
Selection effects on this correlation have also been investigated with contrasting
results \citep{Band05,Ghirlanda05a,Butler07,Ghirlanda08,Butler09,Shahmoradi09}.

In this article, the \epeiso{} correlation and 
other spectral energy correlations derived from it confront the most energetic 
GRBs yet detected, GRB\,080916C \citep{Greiner09,Abdo09} and GRB\,090323 \cite{Derhorst09,Golenetskii09b}.
In particular, GRB\,080916C, with its
huge energy release, with the extension of the spectrum of its prompt emission
up to tens of GeV without any excess or cut--off, and with
the accurate measurements of its spectral parameters provided by \fermi/GBM and \konus/WIND, 
gives us a unique opportunity of further testing the robustness and extension
of these correlations and investigating their properties.
We also update the \epeiso{} correlation by including 
the new detected GRBs with known redshift and \epi, and compare the best estimate of its
normalization as obtained
by using only GRBs detected by \fermi/GBM with those estimated with 
other instruments.
Our study is based on published spectral 
results by \konus/WIND, \fermi/GBM, \swift{} and on specific data 
analysis of publicly available data. 
   \begin{figure}
  \includegraphics[width=\columnwidth]{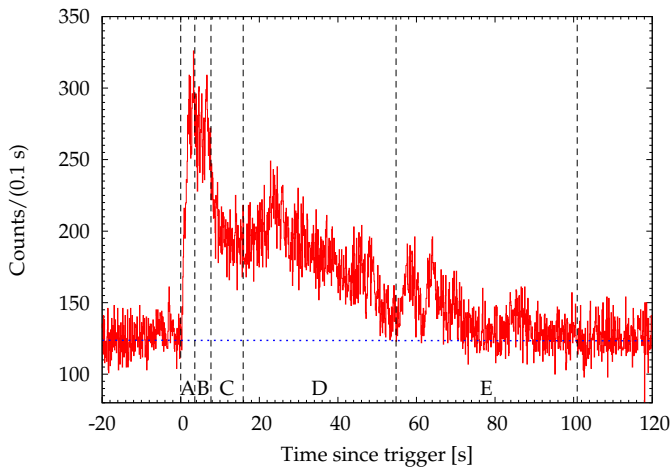}
      \caption{Light curve of the prompt emission of GRB\,080916C as measured by the \fermi/GBM - n3 detector ($\sim$8--1000 keV). 
      The horizontal dotted line is the best--fit of the 
      background level as measured before and after the GRB. Also shown (vertical dashed lines) are the time intervals for which time--resolved spectra from 8 keV up to 10 GeV have been reported by Abdo et al. (2009).
              }
         \label{Fig2}
   \end{figure}
   \begin{figure}
%  \hspace{0.5cm} \includegraphics[width=6cm,angle=-90]{080916C_XRT_both.eps}
  \includegraphics[width=\columnwidth]{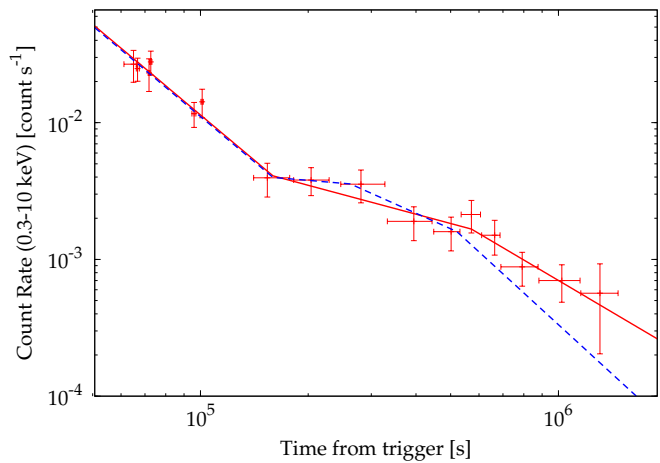}
      \caption{X--ray afterglow light curve of GRB\,080916C as measured by the \swift/XRT in 0.3--10 keV. The continuous line shows the best--fit double broken power--law; the dashed line shows the triple broken power--law obtained by fixing the last slope to 2.4 and corresponding to the 90\% c.l. lower limit to \tb (see text). 
              }
         \label{Fig3}
   \end{figure}
%__________________________________________________________________

\section{Observations and data analysis}

GRB\,080916C was detected by the \fermi/GBM on 16 September 2008
at 00:12:45 UT as a long, multi--peak structured GRB with duration
$T_{90}$ $\sim$66~s
in 50 -- 300 keV \citep{Goldstein08}. The light curve of the prompt
emission as measured by one of the \fermi/GBM NaI detectors which triggered the 
event is shown in Fig. 1. The burst was
observed also by AGILE (MCAL, SuperAGILE, and ACS), RHESSI, INTEGRAL (SPI--ACS), 
\konus/Wind and MESSENGER \citep{Hurley08}. Remarkably, very high energy photons
from GRB\,080916C were detected by \fermi{}/LAT up to $\sim$10 GeV, with more than 
145 photons above 100 MeV and 14 photons above 1 GeV \citep{Tajima08,Abdo09}.

Thanks to the prompt dissemination of the \fermi{}/LAT and IPN 
positions, 
GRB\,080916C was followed--up by \swift{} and other ground telescopes, 
leading to the 
detection of both the X--ray and optical fading counterparts. Of particular 
interest are the afterglow measurements by \swift/XRT and GROND. The X--ray 
afterglow light curve (Fig. 2) 
shows the canonical shape: a steep decay followed by a flat 
decay and than a steeper power--law decay with index $\sim$1.4 and no break up to 
$\sim$1.3 Ms from the GRB onset \citep{Stratta08}. The optical afterglow light 
curve shows a different behavior, with strong evidence of a simple power--law decay 
\citep{Greiner09}.

Later on, 
a photometric redshift of 4.35$\pm$0.15 was reported 
by the GROND team \citep{Greiner09}. 
By combining this value with the fluence and spectral 
parameters of 
the prompt emission
provided by \fermi/GBM and \konus/WIND, 
080916C was found to be the most energetic GRB ever, 
with an \eiso{} of $\sim$4$\times$10$^{54}$ erg in the standard 
1--10000 keV cosmological 
rest--frame energy band. Moreover, the joint 
spectral analysis of \fermi/GBM and LAT data published by the \fermi{} 
team \citep{Abdo09}
showed 
that the spectrum extends 
up to 
$\sim$1--10 GeV without any excess or cut--off 
and that the \eiso{} computed
by integrating up to 10 GeV 
is as huge as $\sim$10$^{55}$ erg. 

More recently, another very bright GRB\,090323  has been detected and localized
by the \fermi{}/LAT \citep{Ohno09a}. Follow--up observations were
performed by \swift\
and other ground facilities, leading to the discovery of the X--ray,
optical and radio counterparts \citep{Kennea09,Updike09,Harrison09}. 
A spectroscopic redshift of 3.57 was
measured by Gemini south \citep{Chornock09}. Spectral parameters
and fluence for GRB\,090323 were 
provided by both the \fermi{}/GBM \citep{Derhorst09} 
and \konus/WIND \citep{Golenetskii09b}.
Based on the spectrum and fluence measured by \konus/WIND
and the redshift of 3.57 measured by Gemini south,
it can be found that the \eiso{} value of this event
in the 1--10000 keV cosmological
rest--frame energy band is $\sim$4$\times$10$^{54}$ erg, thus
comparable to that of GRB\,080916C.
No refined analysis of the VHE emission measured by the LAT
from this event has been published.

In our analysis we used results published in the above references and
specific data analysis of the public \fermi/GBM and \swift/XRT data\footnote{The \fermi/GBM
data and analysis tools are
available at {\it ftp://legacy.gsfc.nasa.gov/fermi/};
the \swift/XRT data are
available at {\it http://swift.gsfc.nasa.gov/docs/swift/archive/}}.
In particular, for GRB\,080916C we extracted the light curve
of each \fermi/GBM detection unit which triggered the event (n3, n4 and b0) 
by using the {\it gtbin} tool included in the data reduction and 
analysis tools\footnote{available at 
{\it http://fermi.gsfc.nasa.gov/ssc/data/analysis/}}. The \swift/XRT data 
of this burst were 
processed using the heasoft package
(v.6.4). We ran the task xrtpipeline (v.0.11.6) applying calibration
and standard filtering and screening criteria\footnote{see {\it http://swift.gsfc.nasa.gov/docs/swift/analysis/}}. 

Radiated energies and luminosities are computed
by assuming a standard $\Lambda$CDM cosmology with \h0 = 70 km s$^{-1}$ Mpc$^{-1}$, \omegam = 0.27 and
\omegal = 0.73. The quoted uncertainties are at 68\% c.l., unless differently stated.
\begin{table}
\begin{minipage}[t]{\columnwidth}
\caption{The 25 GRBs with known redshift and measured \epi{} added to the sample of 
Amati et al. (2008) in our analysis of the \epeiso{} correlation, resulting in a 
total
of 95 GRBs. 
}
\label{catalog}
%\centering
\renewcommand{\footnoterule}{}  % to avoid a line before footnotes
\begin{tabular}{lccccc}
\hline \hline
GRB & $z$\footnote{Taken from the GRB Table by J. Greiner and references therein
(http://www.mpe.mpg.de/jcg/grbgen.html).} &  \epi  & \eiso\footnote{Computed
in the 1--10000 keV cosmological rest--frame by assuming a standard $\Lambda$CDM cosmology with \h0 = 70 km s$^{-1}$ Mpc$^{-1}$, \omegam = 0.27 and
\omegal = 0.73.} & Instrument\footnote{Instrument(s) that provided 
the spectral parameters and fluence used for the computation of \epi{} and \eiso: 
HET = HETE--2; BAT = \swift/BAT; KW = \konus/WIND; WAM = \suz/WAM; 
GBM = \fermi/GBM.} & 
Ref.\footnote{References for spectral parameters and fluence: (1) Sakamoto et 
al. (2005)\nocite{Sakamoto05}; 
(2) Golenetskii et al. (2007)\nocite{Golenetskii07};
(3) Ohno et al. (2008)\nocite{Ohno08}; (4) Barthelmy et al. (2008a)\nocite{Barthelmy08a};
(5) Golenetskii et al. (2008a)\nocite{Golenetskii08a}; (6) Golenetskii et al. (2008b)\nocite{Golenetskii08b}; (7) Golenetskii et al. (2008c)\nocite{Golenetskii08c};
(8) Golenetskii et al. (2008d)\nocite{Golenetskii08d};
(9) Golenetskii et al. (2008e)\nocite{Golenetskii08e};
(10) Meegan et al. (2008)\nocite{Meegan08};
(11) Pal\'shin et al. (2008)\nocite{Palshin08};
(12) Bissaldi et al. (2008a)\nocite{Bissaldi08a} and
Baumgartner et al. (2008)\nocite{Baumgartner08};
(13) Bissaldi et al. (2008b)\nocite{Bissaldi08b};
(14) Palmer et al. (2008a)\nocite{Palmer08a};
(15) Barthelmy et al. (2008b)\nocite{Barthelmy08b};
(16) Palmer et al. (2008b)\nocite{Palmer08b} and 
Bhat et al. (2008)\nocite{Bhat08};
(17) Golenetskii et al. (2008g)\nocite{Golenetskii08g};
(18) Bissaldi \& McBreen (2008)\nocite{Bissaldi08c};
(19) Golenetskii et al. (2009a)\nocite{Golenetskii09a};
(20) Golenetskii et al. (2009c)\nocite{Golenetskii09c};
(21) Pal\'shin et al. (2009)\nocite{Palshin09};
(22) von Kienlin (2009)\nocite{Kienlin09};
(23) Connaughton (2009)\nocite{Connaughton09};
(24) see text.
} \\
 & & [\rm keV] & [{\rm 10$^{52}$ erg}] \\
\hline
  020127  & 1.9  & 290$\pm$100  &  3.5$\pm$0.1  & HET & (1)  \\
  071003  & 1.604  & 2077$\pm$286  &  36$\pm$4  & KW & (2)  \\
  080413  & 2.433  & 584$\pm$180  &  8.1$\pm$2.0  & BAT/WAM & (3)  \\
  080413B & 1.10  & 150$\pm$30  &  2.4$\pm$0.3  & BAT & (4)  \\
  080514B & 1.8  & 627$\pm$65  &  17$\pm$4  & KW & (5)  \\
  080603B & 2.69   & 376$\pm$100  &  11$\pm$1  & KW & (6)  \\
  080605  & 1.6398  & 650$\pm$55  &  24$\pm$2  & KW & (7)  \\
  080607  & 3.036  & 1691$\pm$226  &  188$\pm$10  & KW & (8)  \\
  080721  & 2.591  & 1741$\pm$227  &  126$\pm$22  & KW & (9)  \\
  080810  & 3.35  & 1470$\pm$180  &  45$\pm$5  & GBM & (10)  \\
  080913  & 6.695  & 710$\pm$350  &  8.6$\pm$2.5  & BAT/KW & (11)  \\
  080916  & 0.689  & 184$\pm$18  &  1.0$\pm$0.1  & BAT/GBM\footnote{
  Spectral parameters from \fermi{}/GBM and fluence from \swift/BAT.} & (12)  \\
  081007  & 0.5295  & 61$\pm$15 & 0.16$\pm$0.03 & GBM & (13)  \\
  081008  & 1.9685  & 261$\pm$52  &  9.5$\pm$0.9  & BAT & (14)  \\
  081028  & 3.038  & 234$\pm$93  &  17$\pm$2  & BAT & (15)  \\
  081118  & 2.58  & 147$\pm$14  &  4.3$\pm$0.9  & BAT/GBM$^{\it e}$
   & (16)  \\
  081121  & 2.512  & 871$\pm$123  &  26$\pm$5  & KW & (17)  \\
  081222  & 2.77  & 505$\pm$34  &  30$\pm$3  & GBM & (18)  \\
  090102  & 1.547  & 1149$\pm$166 & 22$\pm$4 & KW & (19)  \\
  090328  & 0.736  & 1028$\pm$312 & 13$\pm$3 & KW & (20)  \\
  090418  & 1.608  & 1567$\pm$384 & 16$\pm$4  & BAT/KW & (21)  \\
  090423  & 8.1   & 491$\pm$200 & 11$\pm$3 & GBM & (22)  \\
  090424  & 0.544  & 273$\pm$50  &  4.6$\pm$0.9  & GBM & (23)  \\
\hline
  080916C  & 4.35  & 2646$\pm$566 &380$\pm$80 & GBM/KW & (24)  \\
  090323  & 3.57  & 1901$\pm$343 & 410$\pm$50 & KW & (24)  \\
\hline
\end{tabular}
\end{minipage}
\end{table}

\section{The \epeiso{} correlation: update and comparison among  
different instruments.}

In Fig.~3 (right panel) we show the \epeiso{} correlation for long GRBs (short
GRBs and the peculiar sub--energetic GRB\,980425 are not included) obtained
by adding to the sample of 70 events of Amati et al. (2008)\nocite{Amati08}  
25 more GRBs for which measurements of the redshift and/or of the
spectral parameters have become available in the meanwhile (as of April 2009). 
As in the previous evaluations, \eiso{} is derived in the 1--10000 keV energy
band.
The \epi{} and \eiso{} of these events, together with the redshift and the relevant
references, are reported in Tab.~1. These values and their uncertainties
were computed based on published spectral parameters and fluences and following
the methods and criteria reported, e.g., in Amati (2006)\nocite{Amati06} and
Amati et al. (2008)\nocite{Amati08}. As can be seen, this updated sample of 95 GRBs
is fully consistent with the \epeiso{} correlation and its dispersion as derived
by Amati et al. (2008)\nocite{Amati08}. 
This is quantitatively confirmed by the fit with both
the classical $\chisq$ method and with the maximum likelihood method adopted, e.g., by 
Amati (2006) and Amati et al. (2008)\nocite{Amati06,Amati08}, 
which allows us to quantify the extrinsic scatter of the correlation (\sext). 
We obtain a 
slope $m$ = 0.57$\pm$0.01 and a $\chisq$ of 594, by means of a linear fit 
to the log(\epi) vs. log(\eiso) data points with the 
$\chisq$ method, and $m$ = 
0.54$\pm$0.03 and \sext = 0.18$\pm$0.02 (68\% c.l.)
with the maximum likelihood method. 
These values are fully consistent with those obtained by Amati et al. (2008)\nocite{Amati08}.

Butler et al. (2009)\nocite{Butler09}, in their study of the selection effects, claim 
that the dispersion and significance of the \epeiso{} correlation in the intrinsic plane 
is comparable to that of the \epo -- fluence in the observer plane, and that 
the normalization of the 
\epeiso{} correlation depends on the instrument used to detect GRBs. 
In Fig.~3 (left panel), we show the
distribution of these 95 GRBs in the \epo{} -- fluence observer plane. In order to allow a
reliable comparison, the X and Y scales of this plot cover the same orders of
magnitude as the \epeiso{} plane shown in Fig.~3 (right panel). As can be seen,
when we move from the observer to the intrinsic plane the dispersion of
the correlation between spectral peak photon energy and fluence (radiated energy)
decreases significantly (\sext{} from $\sim$0.31 to $\sim$0.18 and
$\chisq$ from 3110 to 594), its extension covers more orders of magnitudes, and
its significance increases (Spearman's $\rho$ from $\sim$0.75 to $\sim$0.88). 
In Fig.~4 we compare the normalization of the \epeiso{} correlation obtained with
all the most relevant instruments with that derived by Amati et al. (2008)\nocite{Amati08}.
As can be seen, no significant (i.e. above $\sim$1$\sigma$) 
change is found. In particular,  the \epeiso{} correlation 
derived using the \swift{} GRBs in which, unlike those considered by 
Butler et al. (2009)\nocite{Butler09},
\epi{} is really measured with BAT (from the
official \swift{} team catalog by Sakamoto et al. 2009\nocite{Sakamoto08} and/or GCNs) or with 
broad band instruments (mainly \konus/WIND), is fully consistent with
that determined with other instruments.
The \epeiso{} correlation derived from GRBs detected with \fermi/GBM  
is fully consistent with the correlation as determined from other instruments
with narrower energy ranges. Given the unprecedented broad energy coverage of the GBM 
(from $\sim$8 keV up to more than 30 MeV), the derived \epeiso{} correlation is 
certainly not affected by biases in the estimate of 
the spectral parameters (the so called "data truncation" effect, see. e.g., 
Lloyd et al. 2000\nocite{Lloyd00}).

   \begin{figure*}
   \centerline{
  \hspace{-0.5cm}
  \includegraphics[width=1.15\columnwidth]{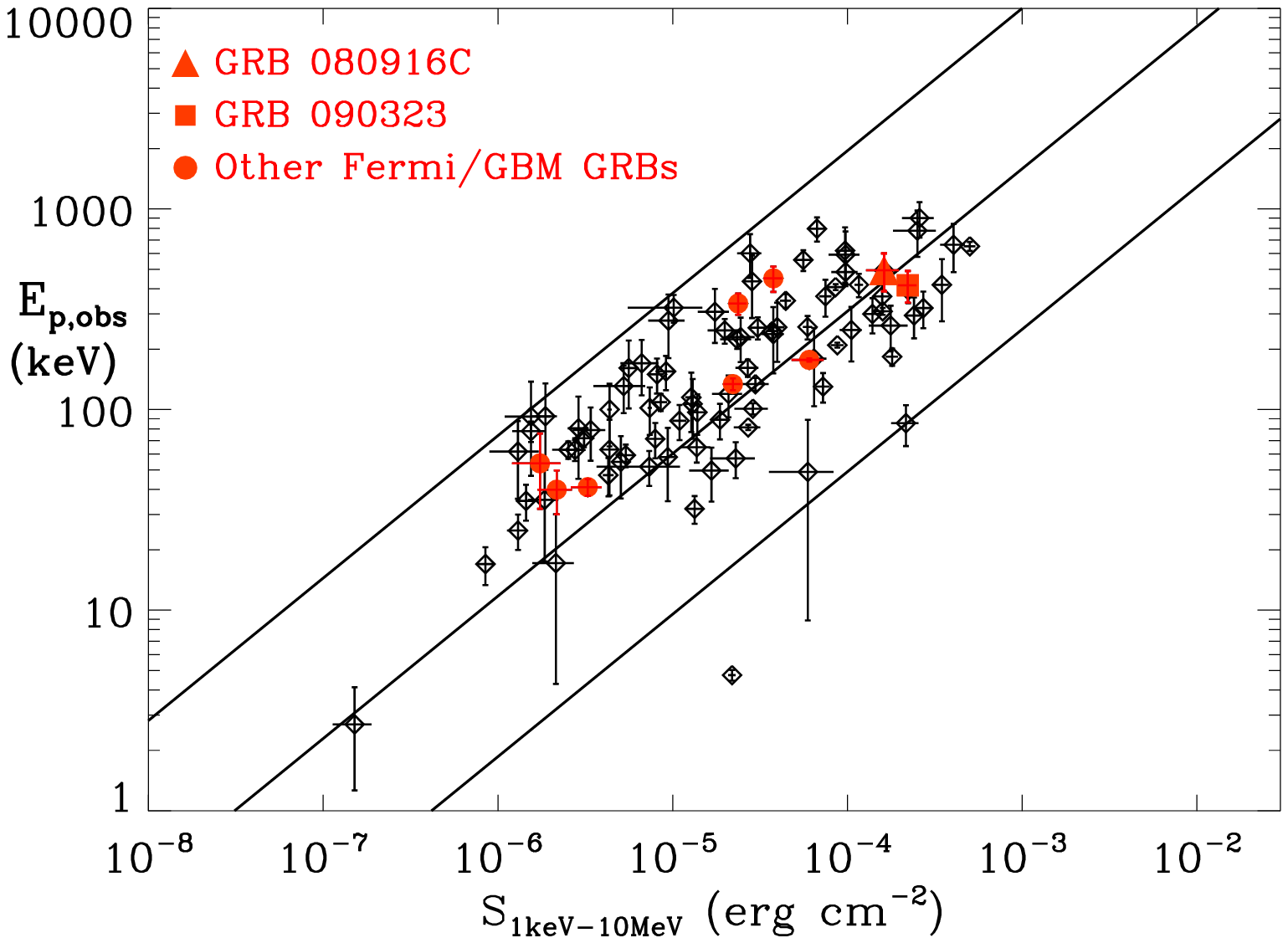}
  \hspace{-1.0cm}
  \includegraphics[width=1.15\columnwidth]{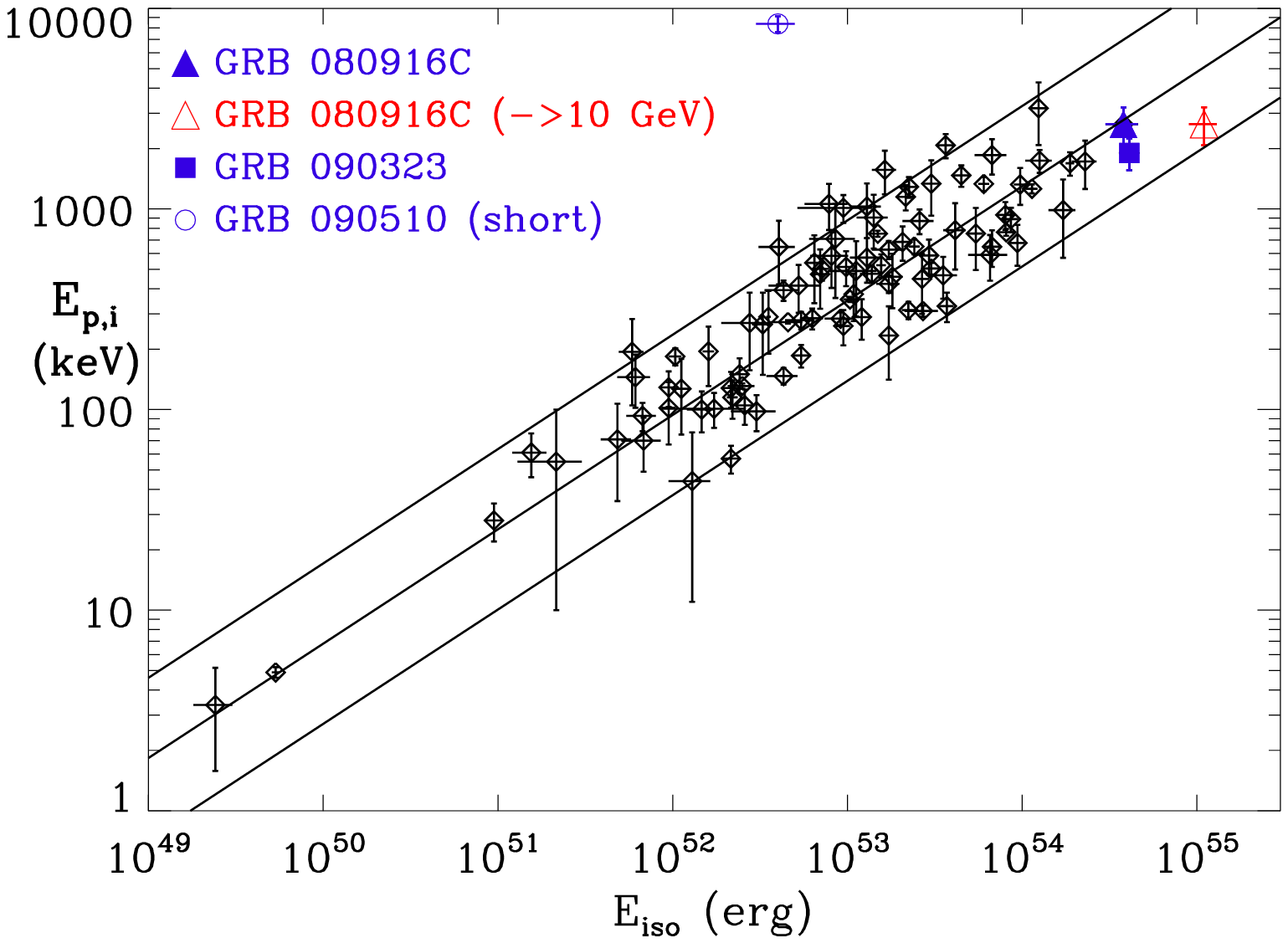}
   }
      \caption{Location in the \epo{} -- fluence (left) and \epeiso{} (right) 
      planes of the 95 GRBs
      with firm redshift and \epo{} estimates as of April 2009 (see text).
      In both panels the points corresponding to the extremely energetic GRBs 
      080916C and 090323 are highlighted. In addition, in the left panel we mark
      with red dots those GRBs with spectral parameters and fluence provided by the \fermi{}/GBM, and in the right panel we also show the GRB\,080916C point 
      obtained with \eiso{} computed 
in the 1 keV -- 10 GeV
cosmological rest--frame energy band and the point corresponding to 
the short GRB\,090510. 
The continuous lines in the right panel correspond to
the best--fit power--law and the $\pm$2$\sigma$
dispersion region of the \epeiso{} correlation as derived by Amati et al. 2008.
              }
         \label{Fig1}
   \end{figure*}

We also tested the effect of the redshift on the \epo{} vs. fluence 
dependence.
Starting from the \epo{} vs. fluence data, we derived 10000 \epeiso{} simulated 
correlations by randomly exchanging the $z$ values among the 95 GRBs, and we 
computed for each sample the Spearman's correlation coefficient $\rho$ between 
the \epi{} and \eiso{} values so obtained. 
We found a $\rho$ distribution fully consistent with a Gaussian with centroid 
$\sim$0.75, 
which is exactly the value obtained for the \epo{} vs. fluence correlation 
in the observer plane, with $\sigma$$\sim$0.035 and extending up to $\sim$0.85.
For comparison, the $\rho$ value of the true  \epeiso{} correlation, as we have seen,
is $\sim$0.88 which is  $\sim$3.8 $\sigma$ from that obtained from the simulation and
corresponds to a chance probability less than 1 over 1000, that the true
\epeiso{} correlation is randomly extracted from the simulated ones.

An exhaustive paper devoted to the discussion of the selection effects
on the \epeiso{} correlation is in preparation.

   \begin{figure*}
   \centerline{
  \hspace{-0.5cm}
  \includegraphics[width=1.15\columnwidth]{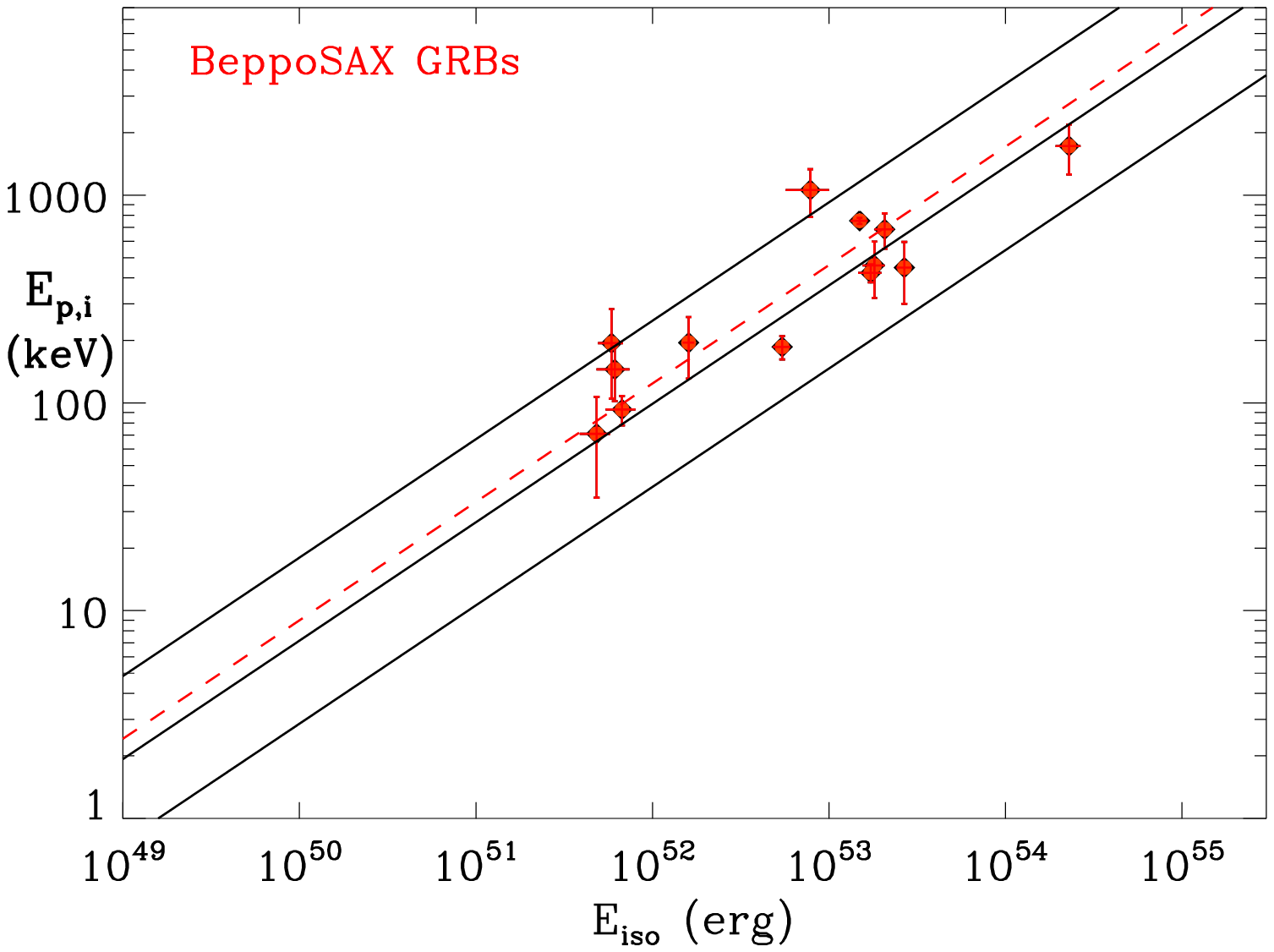}
  \hspace{-1.0cm}
   \includegraphics[width=1.15\columnwidth]{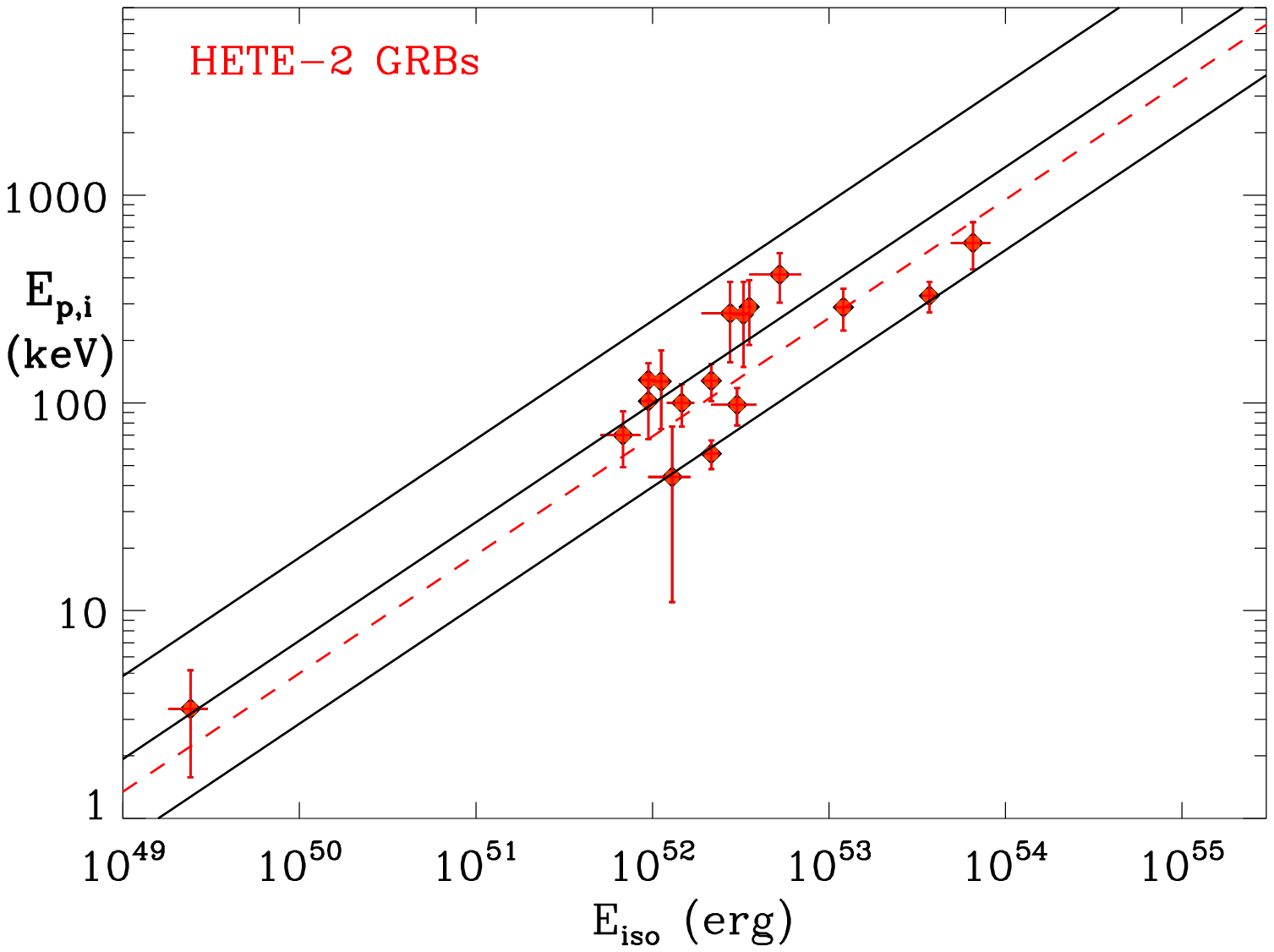}
   }
   \centerline{
  \hspace{-0.5cm}
  \includegraphics[width=1.15\columnwidth]{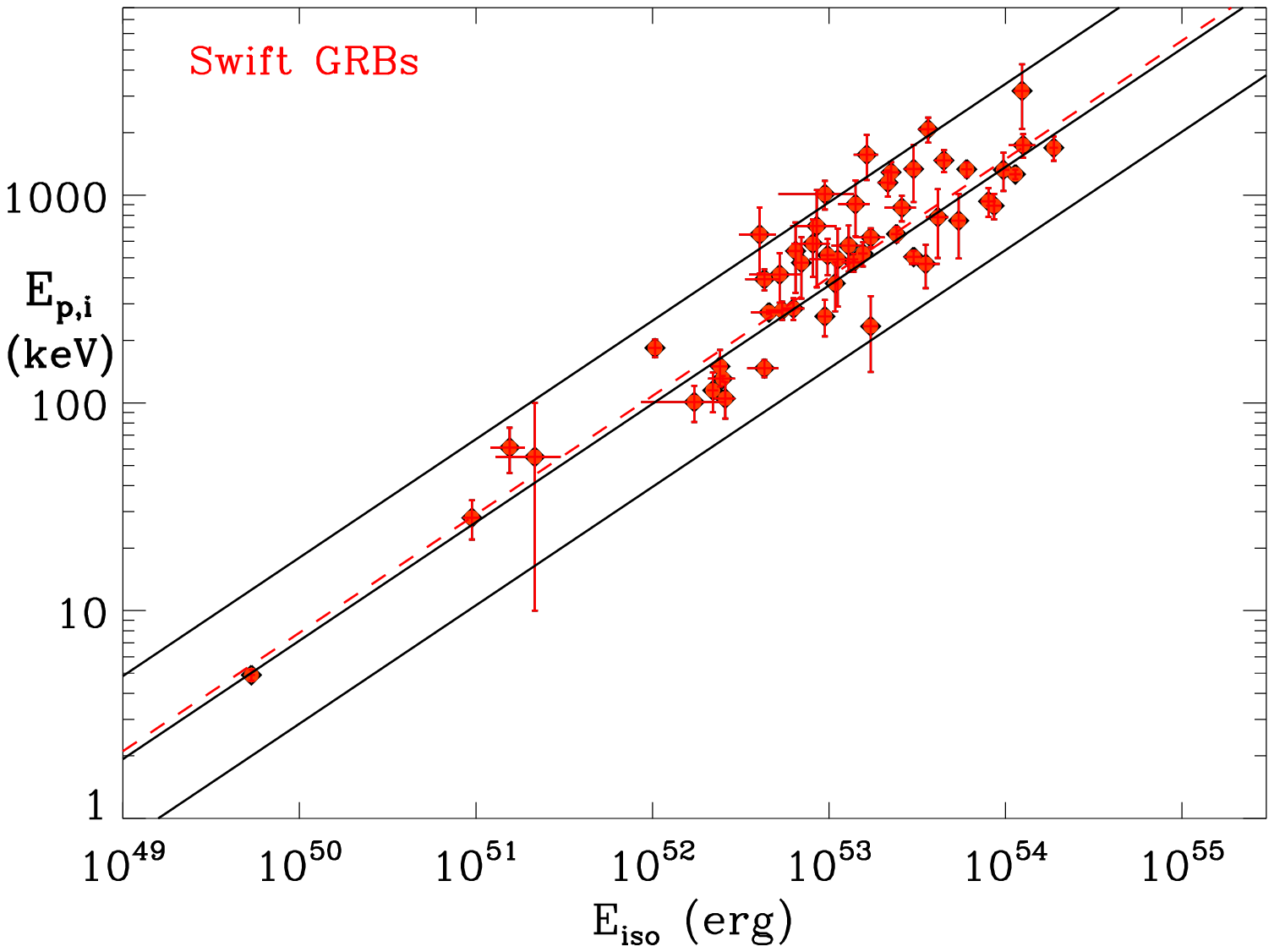}
  \hspace{-1.0cm}
   \includegraphics[width=1.15\columnwidth]{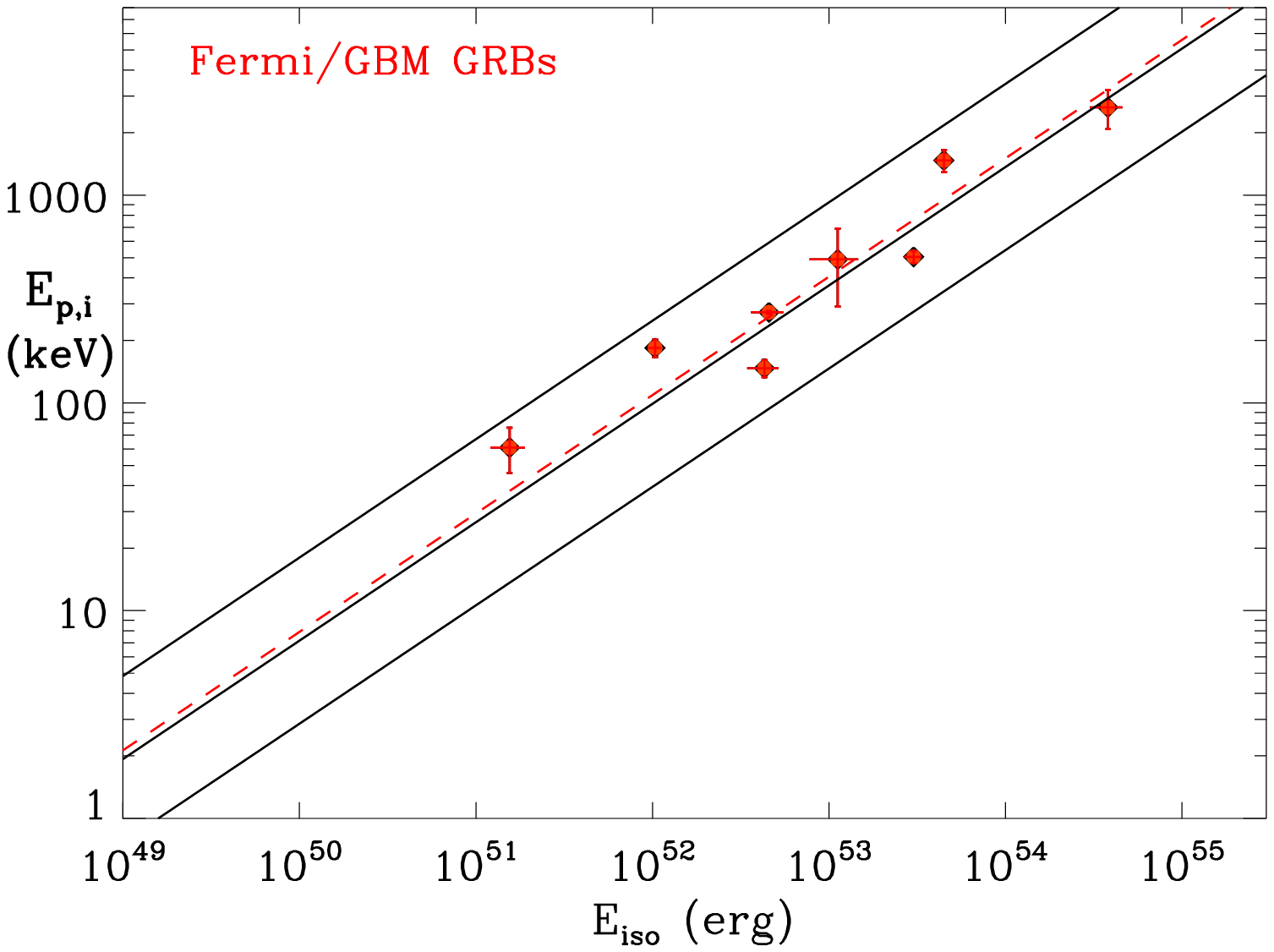}
   }
      \caption{Location in the \epeiso{} plane of those GRBs with 
      localization and \epi{} 
      provided by different instruments. The top panels show those GRBs whose
      detection, localization and spectrum were provided by \sax (left) and 
      HETE--2 (right). The bottom--left panel shows those GRBs
      detected and localized
      by \swift/BAT and for which \epi{} has been provided either by BAT itself or 
      by other 
      instruments (excluding the \fermi/GBM). 
      The bottom--right panel shows those GRBs for which the localization
      has been provided either by \swift{} or \fermi/LAT and \epi{} 
      has been measured
      by \fermi/GBM (right panel). In all panels, the 
continuous lines correspond to
the best--fit power--law and the $\pm$2$\sigma$
dispersion region of the correlation as computed by including all 95 GRBs with
known $z$ and \epi{} and the dashed line is the best--fit power--law obtained
by considering the plotted points only.
              }
         \label{Fig1}
   \end{figure*}

\section{\fermi{} highly energetic GRBs in the \epeiso{} plane.}

Based on \fermi/GBM, the 
fluence of GRB\,080916C in 8 keV -- 30 MeV was $\sim$1.9$\times$10$^{-4}$ erg cm$^{-2}$
and its 
time--averaged spectrum in the same energy band can be fit with a Band function
\citep{Band93} with $\alpha$ = $-$0.91$\pm$0.02, $\beta$ = $-$2.08$\pm$0.06 and
\epo = 424$\pm$24 keV \citep{Derhorst08}. The \konus/WIND team reported, for the
20 keV -- 10 MeV energy band, 
a fluence of (1.24$\pm$0.17)$\times$10$^{-4}$ erg cm$^{-2}$, a 256--ms peak flux of 
(1.19$\pm$0.30)$\times$10$^{-5}$ erg cm$^{-2}$ s$^{-1}$ and a time--averaged 
spectrum with $\alpha$ = $-$1.04$\pm$0.06, $\beta$ = $-$2.26$\pm$0.3 and
\epo = 505$\pm$75 keV \citep{Golenetskii08f}.

By taking into account 
these fluences and spectral parameters, with their uncertainties,
the redshift, with its uncertainty, provided by GROND,
and by integrating the cosmological rest--frame spectrum in the 
commonly adopted 1 keV -- 10 MeV energy band \citep{Amati02,Amati06}, we derive the 
following values: \eiso = (3.8$\pm$0.8)$\times$10$^{54}$ erg, \epi = 2646$\pm$566
keV. 
As can be seen 
in Fig.~3, with these values the location of GRB\,080916C in the \epeiso{} plane
is very close to the best--fit power--law obtained with
the sample of 
70 long GRBs considered by Amati et al. (2008)\nocite{Amati08}. 
This confirms that GRB\,080916C follows 
the \epeiso{} correlation and extends its range of validity along \eiso{} by a factor 
of $\sim$2. 
If GRB\,080916C is excluded from the fit of
the correlation, the values of the parameters and their uncertainties do not change
significantly with respect to those reported in the previous Section, which is the
case when the softest/weakest events are excluded.
This confirms 
that the significance and characterization of the \epeiso{} 
correlation do not depend
on events at the extremes of the ranges of \epi{} and \eiso.

The spectral analysis performed by Abdo et al. (2009), shows that the time 
resolved spectra of this event can be fit with the simple Band function from 
$\sim$8 keV up to more than 1 GeV. This implies that the \eiso{} above 10 MeV could be not negligible.
Indeed, by extending the integration up to 10 GeV 
(cosmological rest--frame), and using the $\beta$ value provided by \fermi/GBM 
(which, given the extension of the energy band of this instrument, is expected to 
be more accurate than that provided by \konus/WIND), we obtain a value 
of \eiso{} of (1.1$\pm$0.2)$\times$10$^{55}$ erg, which is higher by a factor of 
$\sim$2.5. As can be seen in Fig. 3, with this (huge) value of 
\eiso{}, GRB\,080916C 
is still consistent with the \epeiso{} correlation within 2$\sigma$ and extends its 
dynamic range along \eiso{} by about half an order of magnitude. 

In Fig.~3 we also show the location in the \epeiso{} plane of the other 
ultra--energetic GRB detected more recently by \fermi, GRB\,090323. For this event,
no refined analysis of the VHE emission measured by the LAT 
has been published, thus no reliable
extrapolation and integration of the spectrum up to the GeV range can be done. Hence,
we restrict the analysis to the standard 1 keV -- 10 MeV energy band. In addition,
the published GBM spectral analysis concerns only the first $\sim$70 s of the event
(which shows a total duration of $\sim$120 s), and thus these data do not provide
a reliable estimate of \epi{} and \eiso.
By using the spectral parameters $\alpha$ = $-$0.96$_{-0.09}^{+0.12}$, 
$\beta$ = $-$2.09$_{-0.22}^{+0.16}$, 
\epo = 416$_{-73}^{+76}$ keV and the fluence of 
(2.0$\pm$0.3)$\times$10$^{-4}$ erg cm$^{-2}$ (20 keV -- 10 MeV) 
provided by 
\konus/WIND \citep{Golenetskii09b},
together with the redshift of 3.57 measured by Gemini south,
we find \eiso = (4.1$\pm$0.5)$\times$10$^{54}$ erg
and \epi = 1901$\pm$343 keV. These values are very close to those of
GRB\,080916C and make also this event fully consistent with the \epeiso{}
correlation. This is further evidence that the newly discovered class of
extremely energetic GRBs follows the correlation.
The detection of more GRBs with
photons at GeV energies (e.g., from \fermi/LAT) will strengthen this result.

The extension of the spectrum of GRB\,080916C supports the possibility that, 
at least for a fraction of 
long GRBs, the commonly adopted 1 keV -- 10 MeV cosmological 
rest--frame energy band for the computation of \eiso{} may lead to an underestimate 
of 
this quantity and be a source of systematics and extra--scatter in the \epeiso{} 
correlation. To test this, we considered again the sample of Amati 
et al. (2008)\nocite{Amati08} plus the 25 GRBs reported in Tab.~1.
For each event, 
we re--computed the \eiso{} value by extending the integration up to 10 
GeV using the $\alpha$ and $\beta$ values reported in the literature. 
For those events without a reported value of $\beta$, e.g. in the 
case of a fit with a cut--off power--law, we adopted
a Band function with 
$\beta = -2.3$. 
The fit with the $\chisq$ method provides $m$ = 0.55$\pm$0.01 
with a best--fit $\chisq$ of 619, while the maximum likelihood method provides
$m$ = 0.51$\pm$0.03 and \sext = 0.18$\pm$0.02 (68\% c.l.).
We conclude that extending the computation of \eiso{} up to 10 GeV slightly 
flattens the slope of the \epeiso{} correlation but does not significantly change 
its 
scatter.

Finally, very recently the \fermi/LAT detected and localized 
GeV emission from a bright short ($\sim$0.5s) GRB\,090510 \cite{Ohno09b}.
This event was also detected by AGILE at energies above
100 MeV \cite{Longo09}. By combining the VLT redshift estimate of $z$=0.903
\cite{Rau09} and the spectral parameters and fluence 
obtained with the \fermi/GBM \cite{Guiriec09}, it results
that the \eiso{} and \epi{} of this event are 
(4$\pm$1)$\times$10$^{52}$ erg
and 8370$\pm$760 keV, respectively. With these values GRB\,090510 lies in the \epeiso{}
plane significantly above the region populated by long GRBs (Fig.~3, right panel),
further confirming that short GRBs do not follow the \epeiso{} 
correlation.

   \begin{figure*}
   \centerline{
  \hspace{-0.5cm}
  \includegraphics[width=1.1\columnwidth]{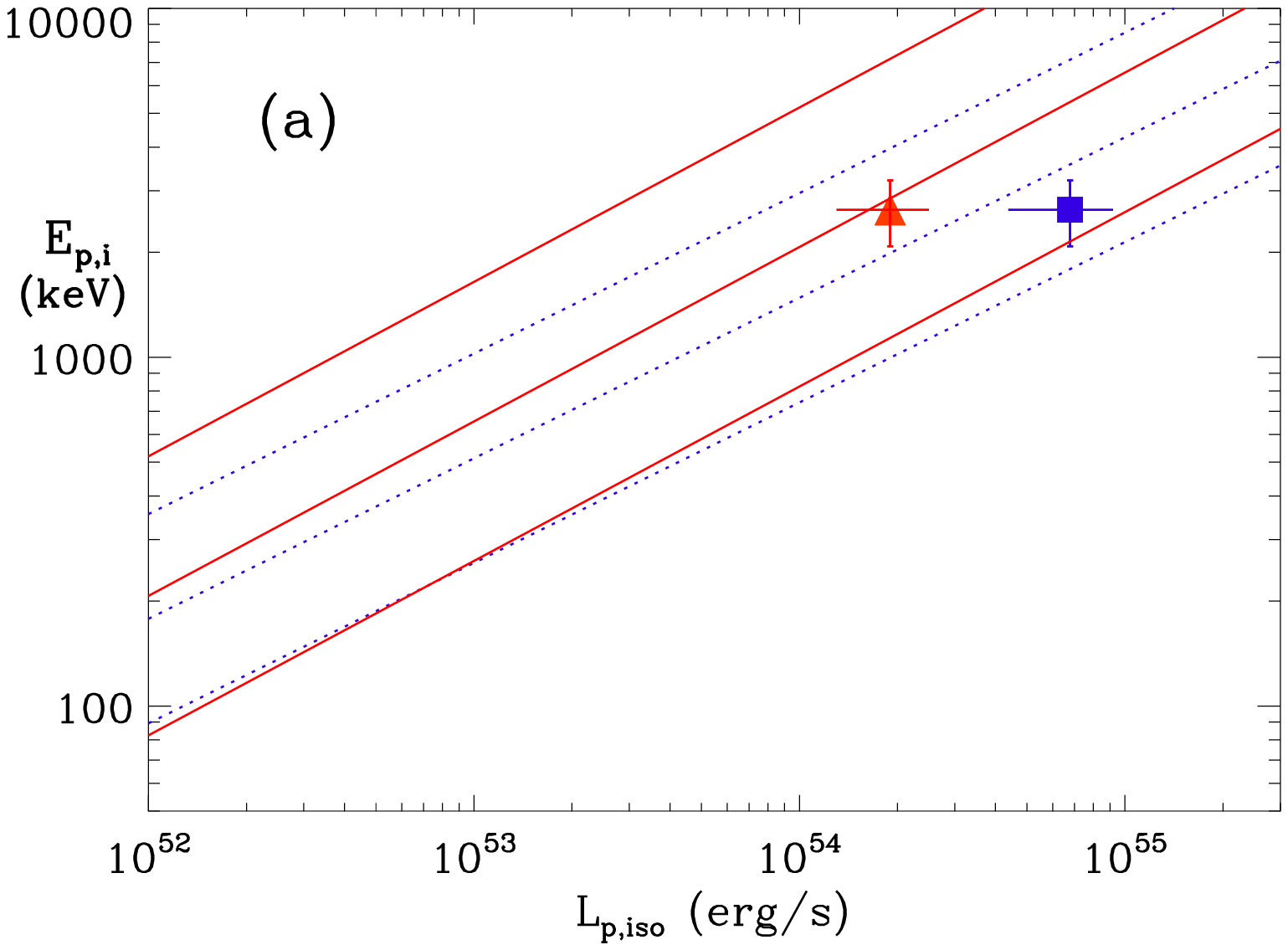}
  \hspace{-1.0cm}
  \includegraphics[width=1.1\columnwidth]{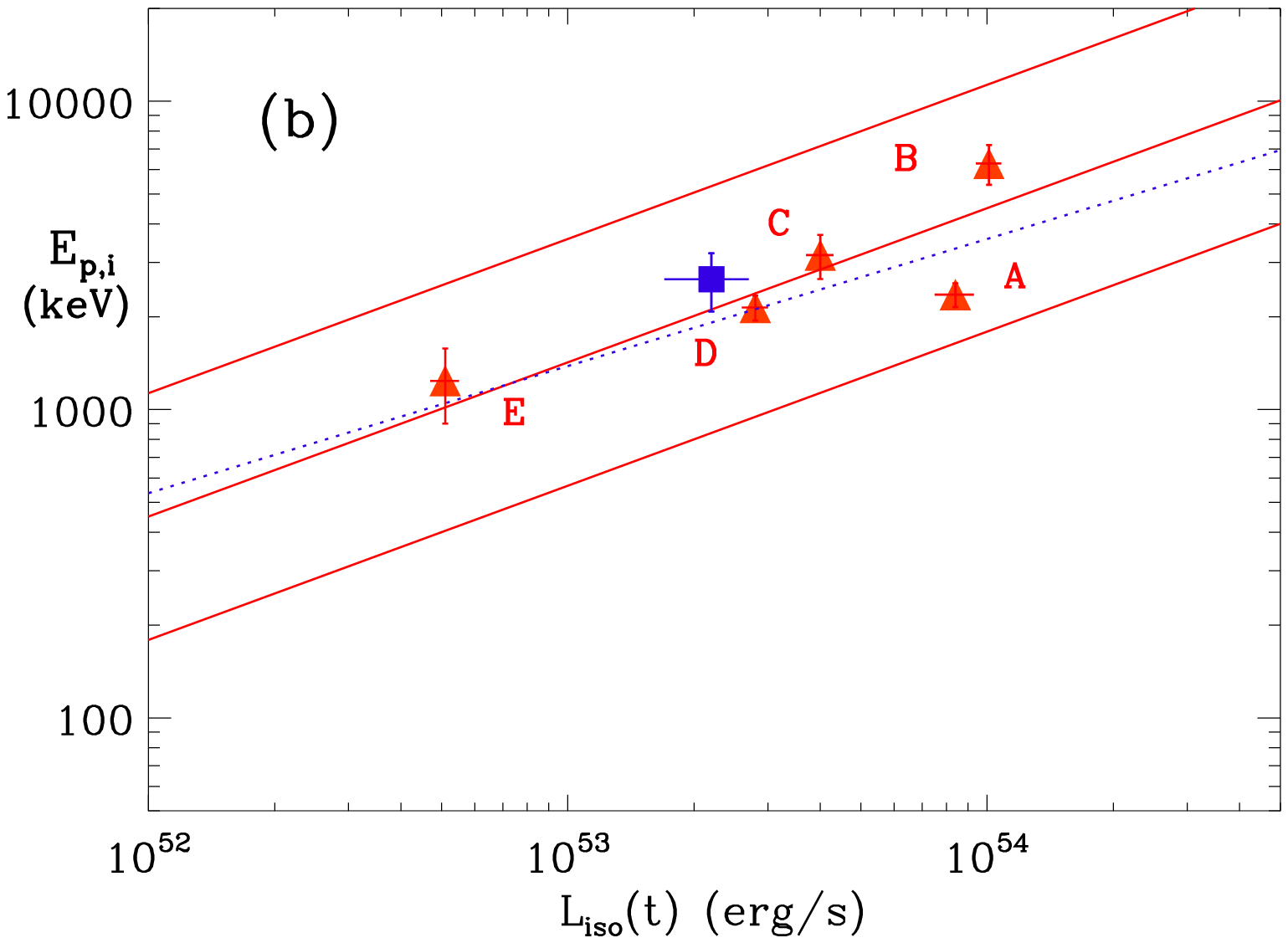}
   }
   \centerline{
  \hspace{-0.5cm}
  \includegraphics[width=1.1\columnwidth]{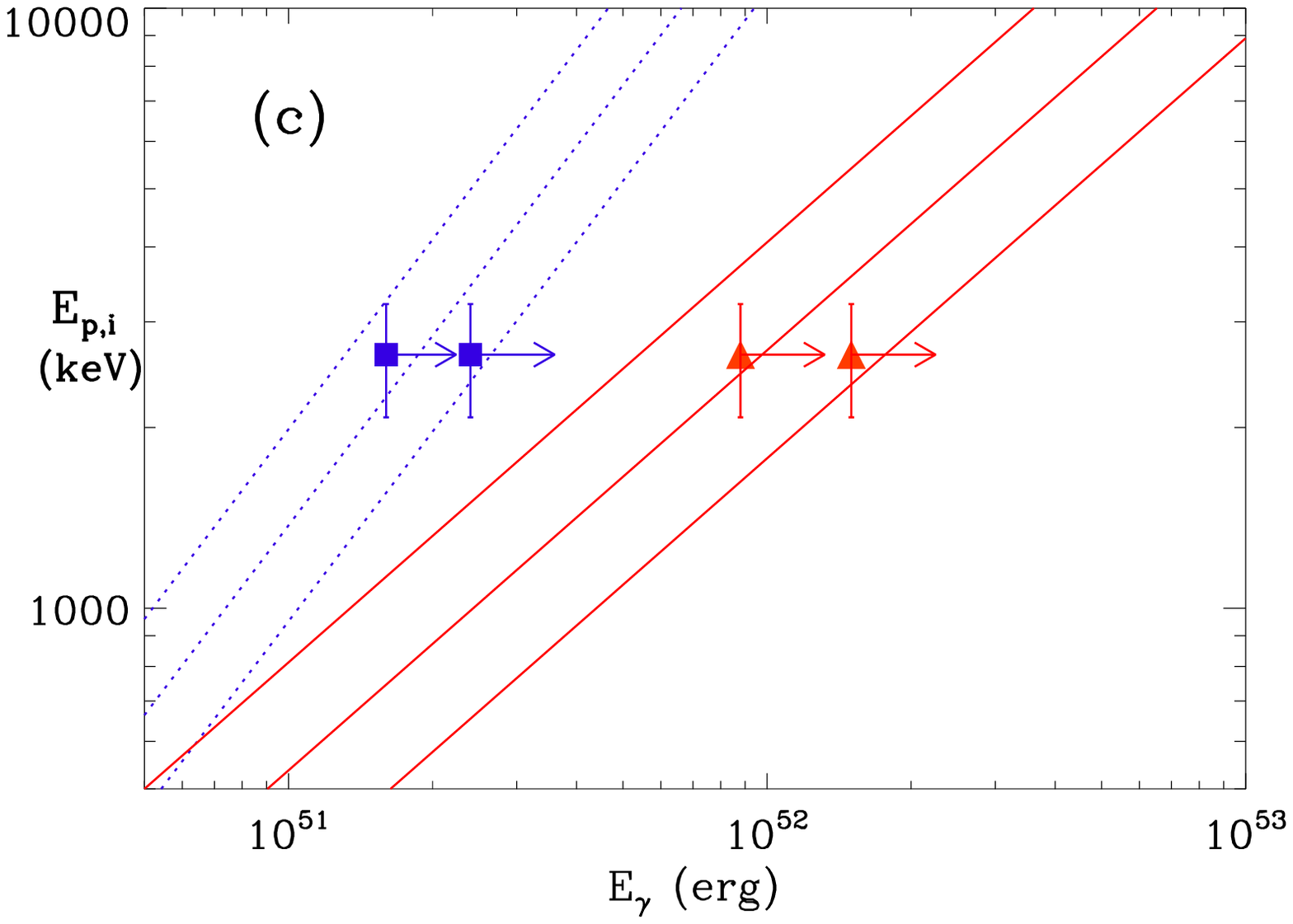}
  \hspace{-1.0cm}
  \includegraphics[width=1.1\columnwidth]{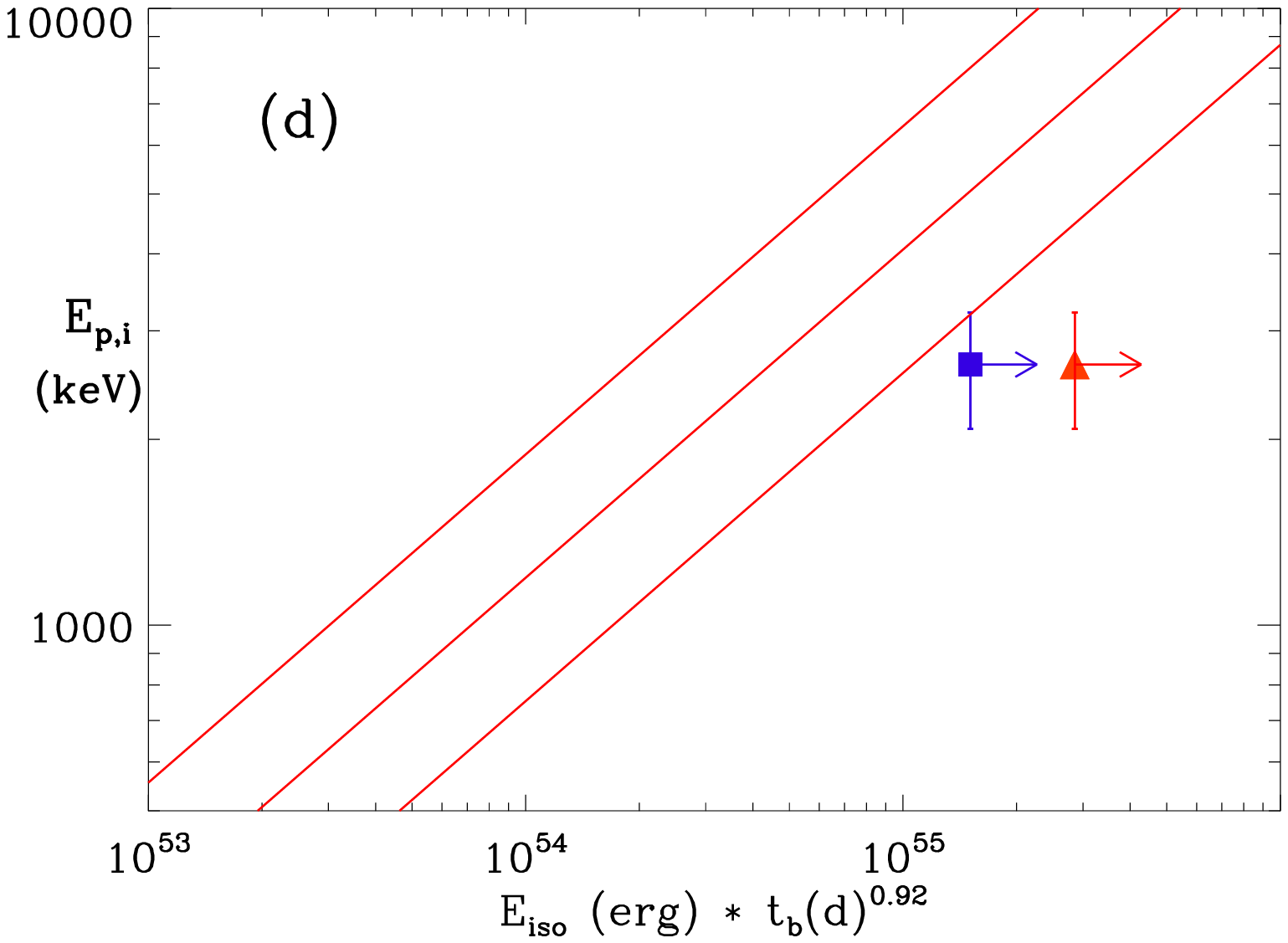}
   }
\caption{
(a) Correlation between \epi{} and \lpiso. 
The red triangle corresponds to GRB\,080916C and the red continuous lines
correspond to the best--fit and 2$\sigma$ range as determined by 
Yonetoku et al. (2004). The blue square is the \lpiso{} of GRB\,080916C 
multiplied by $T_{0.45}^{0.43}$ and the blue dotted lines correspond
to the best--fit and 2$\sigma$ range of the \epi -- \lpiso -- $T_{0.45}$
correlation as determined by Rossi et al. (2008). (b)
Correlation between \epi{} and \liso{} within the GRB (red triangles) based on the 
time resolved spectral analysis reported by Abdo et al. (2009). The letters indicate 
the corresponding time interval (see Fig.~1). The blue square 
corresponds to the luminosity of the whole GRB as determined from the 
time--averaged spectrum. The red continuous lines show the best--fit power--law and the 
2$\sigma$ region of the \epi (t) -- \liso (t) correlation (from Ghirlanda et al. 2007); 
the blue 
dotted line is the best--fit power--law to the 6 points of GRB\,080916C.
(c) Correlation between \epi{} and the collimation--corrected radiated energy 
\ega. The red continuous lines and triangles refer to the homogeneous circum--burst 
medium case while the blue dotted lines and squares refer to the wind case. The lines 
indicate the best--fit power--laws and 2$\sigma$ regions as reported by Ghirlanda et 
al. (2007). For both cases, we report the lower limits to \ega{} corresponding to \tb 
$>$ 1 Ms and \tb $>$ 0.5 Ms (see text). (d) Correlation between \epi, \eiso{} 
and \tb. The lines correspond to the best--fit power--laws and 2$\sigma$ region as 
reported by Ghirlanda et al. (2007). The two lower limits correspond to \tb{}  
$>$ 1 Ms and \tb{}  $>$ 0.5 Ms (see text).
              }
         \label{Fig1}
   \end{figure*}

\section{GRB\,080916C and other spectral energy correlations.}

After the discovery and first studies of the \epeiso{} correlation, it was pointed out 
that \epi{} also correlates with other GRB intensity indicators, like the peak 
luminosity, \lpiso, \citep{Yonetoku04,Ghirlanda05b} or the average 
luminosity, \liso, \citep{Lamb04,Ghirlanda09}. In addition, it was found that 
by including the break time of the afterglow light curve, \tb, 
either directly \citep{Liang05,Nava06,Ghirlanda07} or by using it to derive the 
jet opening 
angle and thus compute the collimation--corrected radiated energy \ega{} 
\citep{Ghirlanda04,Nava06}, the 
extrinsic scatter decreases significantly. As discussed, 
e.g., by Amati (2008)\nocite{Amati08b}, given the strong correlation between \eiso, 
\liso{} and \lpiso, the two--parameters spectral energy correlations are in fact 
equivalent. In addition, in the light of the \swift{} results on X--ray 
afterglow light curves, the 
measurement of \tb{}  and its use to derive the jet opening angle are questioned
\citep{Campana07,Ghirlanda07}. It was also proposed that the inclusion of the "high 
signal time scale", $T_{0.45}$, introduced and used for variability studies, reduces 
the dispersion of the \epeiso{} correlation \citep{Firmani06}, but this property was 
not confirmed by later studies \citep{Rossi08,Collazzi08}. Finally, there is 
evidence that, at least for a significant fraction of GRBs, the correlation 
between \epi{} and luminosity also holds for the time--resolved spectra of individual 
events \citep{Liang04,Firmani08,Frontera09}.

Given its extreme energetics and the good sampling of its optical and X--ray afterglow 
light curve, GRB\,080916C can be used to test
also these correlations. 

Regarding the \epi{}--$L_{iso}$ correlation, from the 256 ms peak flux 
measured by 
\konus{} 
and by assuming the best--fit model of the time averaged spectrum 
we derive \lpiso = (1.9$\pm$0.6)$\times$10$^{54}$ erg s$^{-1}$. 
This value, combined with the \epi{} value of 2646$\pm$566 
keV above derived, gives a data point fully consistent with this correlation (Fig.~5a). 

Regarding the \epi -- \lpiso -- $T_{0.45}$  correlation, from the background subtracted
8--1000 keV light curves obtained with the two (n3 and n4) \fermi/GBM NaI detectors
(see Fig.~1), 
we estimated $T_{0.45}$ (19.5$\pm$0.6 s for $n3$ and 19.2$\pm$0.5 s for $n4$)
following the same approach followed by Rossi et al. (2008)\nocite{Rossi08}. The result is that 
GRB\,080916C is consistent also with this correlation (Fig.~5a).  

From the time--resolved spectral analysis reported by Abdo et al. (2009), an accurate
estimate of \epi\ was obtained.  
By using these results, we computed the 
\liso{} for each of the corresponding time intervals and we reconstructed the track of GRB\,080916C in the 
\epi -- \liso{} plane (Fig.~5b). As can be seen, the spectral
and luminosity evolution of this GRB is fully consistent with the \epi -- \liso{} 
correlation, as typically observed for 
bright events \citep{Liang04,Firmani08,Frontera09}. The slope of the power--law 
that best fits the 6 GRB\,080916C data points is $\sim$0.4, slightly flatter than the
commonly found value of $\sim$0.5. This is mostly due to the data point corresponding to
the first time interval (A) (Fig.~1), which slightly deviates from the 
\epi -- \liso{} correlation. 

Regarding the \epega correlation, no evidence of a jet break \tb\ is found either
in the X--ray light curve (see Fig.~2) up to the end of the XRT observations 
($\sim$1.3 Ms from the trigger) or in the optical light curve up to the end of the 
GROND \citep{Greiner09} observations ($\sim$0.5 Ms from the trigger). 
A 90\% lower limit of about 0.5 Ms to the jet break time is obtained from the X--ray light
curve when assuming a typical post--break slope of 2.4 (see Fig.~2).
By adopting a lower limit of 0.5 Ms to \tb, standard assumptions on 
efficiency of conversion of the fireball kinetic energy into radiated energy, 
on the ISM density and profile, and on the ratio between mass loss rate 
and wind velocity 
\citep{Nava06,Ghirlanda07}, we obtain \ega $>$8.8$\times$10$^{51}$ erg in the
case of an homogeneous circum--burst medium and \ega $>$ 
1.6$\times$10$^{51}$ erg in the case of a wind medium. 
If \tb$>$1~Ms, these values move to \ega $>$1.5$\times$10$^{52}$ erg  
(homogeneous circum--burst medium), \ega $>$2.4$\times$10$^{51}$ erg (wind 
medium).
These lower limits take into account both 
the uncertainty on \eiso{} and on 
the GRB redshift. 
As can be seen in Fig.~5c, the lower 
limits to \ega{} obtained with \tb$\sim$1 Ms lie at around 1.5--2$\sigma$ from the 
best--fit law reported by Ghirlanda et al. (2007)\nocite{Ghirlanda07}, 
while for \tb$\sim$0.5 Ms they 
are fully consistent with it. 

Intriguingly, we find that GRB\,080916C is a possible 
outlier to the \epeisotb{} correlation (Fig.~5d). Indeed, its 
deviation from the best--fit power--law (as determined by Ghirlanda et al. 2007) is 
$>$$\sim$3.5$\sigma$ for \tb $>$ 1Ms and more than $\sim$2.2$\sigma$ for \tb $>$ 0.5Ms. 

\section{Discussion}

The extreme energetics of GRB\,080916C and the fact that its spectrum follows the 
simple Band function without any break or excess up to several tens of GeVs (in the 
cosmological rest--frame) challenge GRB prompt emission models. 
For instance, Abdo et al. (2009)\nocite{Abdo09} and Wang et al. (2009)\nocite{Wang09} suggest that the favored emission 
mechanism is the standard non--thermal synchrotron radiation from 
shock--accelerated electrons within a fireball with bulk Lorentz factor $\Gamma$ 
$>$ $\sim$600 -- 1000 \citep{Abdo09,Greiner09,Li09}. Nevertheless, the lack of a 
synchrotron self--Compton component cannot be explained by 
this scenario and Inverse 
Compton in residual collisions maybe needed to explain time delayed GeV 
photons \citep{Li09}. The fact that GRB\,080916C is fully consistent with the 
\epeiso{} correlation (Sect.~4 and Fig.~3), and with most of the correlations 
derived from it (Fig.~5), further supports the hypothesis that, despite its huge 
isotropic--equivalent radiated energy and the extension to its emission up to VHE, 
the physics behind the emission of this event is not peculiar with respect to 
less energetic long GRBs and XRFs. We note that the 
\epeiso{} correlation itself can be explained within the non--thermal synchrotron 
radiation scenario, e.g., by assuming
that the minimum Lorentz factor, $\gamma$$_{min}$, and the 
normalization of the power--law 
distribution of the radiating electrons do not vary significantly from burst to burst 
or by imposing limits to the slope of the correlation
between the fireball bulk Lorentz factor, $\Gamma$,
and the burst luminosity
\citep{Lloyd00,Zhang02}. 
The consistency of time--resolved spectra of 
GRB\,080916C with the \epi\--luminosity correlation (Fig.~5) confirms that the prompt 
emission is dominated by a single emission mechanism. 
However, the slight
deviation from this correlation of the peak energy and luminosity measured
during the first time interval (Fig.~5--b) may suggest that during the rise
phase of the GRB the main emission mechanism is not still fully at work and other
mechanisms may play a relevant role.

In turn, GRB\,080916C confirms the robustness of the \epeiso{} correlation at least
in the range of intrinsically medium--bright GRBs and, when 
integrating the spectrum up
to 10 GeV, extends it by
$\sim$half an order of magnitude along \eiso{}
(Fig.~3).
The flattening of the slope predicted in some scenarios like, e.g., the multiple subjet model
by Toma et al. (2005) \nocite{Toma05}
or an increase of the dispersion at very high 
energies 
is not observed. 

The above considerations are further supported by  
the other extremely energetic GRB\,090323 
detected more recently by the \fermi{}/LAT,
which shows \epi{} and \eiso{} values similar to those of 
GRB\,080916C and thus is also consistent with the \epeiso{} 
correlation (Fig.~3). 
The recent measurement by \fermi{}
of the \epi{} and GeV emission of the short bright GRB\,090510, combined with
the redshift measurement by VLT, provides further and strong evidence that 
short GRBs do not follow the correlation holding for long ones, 
and that the \epeiso{} plane is a powerful tool 
to discriminate between the two classes and understand their different emission 
mechanisms and origin.

As a part of our study, we have shown (Sect.~3) that: {\it i}) the distribution of the updated
sample of 95 long GRBs with firm estimates of \epi{} and $z$ in the \epeiso{} plane
is fully consistent 
with the slope, normalization and dispersion determined based on previous samples
(Fig.~3);
{\it ii}) moving from the observer frame (\epo{}--fluence) to the intrinsic plane 
(\epi{}--\eiso) the dispersion of the correlation decreases 
and its significance significantly increases (Fig.~3); {\it iii}) if we randomly
re--distribute the redshift values among the
95 GRBs of the sample, the \epi{} vs. \eiso{} distribution is similar to 
to that of \epo{} vs. fluence; 
{\it iv}) not only all \fermi/GBM GRBs but also all the
other long GRBs with known redshift (except GRB\,980425) which have been detected 
with \sax, HETE--2, and \swift, provide \epeiso{} correlations that are  
fully consistent with each other and with 
the \epeiso{} correlation as derived by 
Amati et al. (2008)\nocite{Amati08} (Fig.~4).

All this evidence contrasts the conclusions by Butler et al. 2009
\nocite{Butler09}) that the
\epeiso{} correlation is strongly affected by instrumental effects. 
In addition, the fact that GRBs detected and localized in different energy bands and 
by different instruments all follow the \epeiso{} correlation favour
spectral energy correlations not being strongly affected by selection 
effects introduced in the observational process that leads to the redshift estimate.
An exhaustive analysis of instrumental and selection effects on the \epeiso{} is 
under way and
will be reported elsewhere.

The spectrum of GRB\,080916C following the Band function without any 
cut--off up to a few tens of GeVs (in the cosmological rest--frame) may suggest 
that the commonly adopted 1 keV -- 10 MeV energy band is too narrow for a correct 
computation of \eiso, thus biasing the \epeiso{} correlation. However, our analysis 
reported in Sect.~4 shows that the extension of the energy band up to 10 GeV 
which \eiso{} is 
computed has a marginal impact on the slope and the dispersion of the correlation, 
further supporting its robustness. 

Finally, testing the consistency of very high energy GRBs with the \epeiso{} and other 
spectral energy correlations (Sect.~5) is important for their potential use for 
cosmology \citep{Ghirlanda06,Amati08}. Indeed, due to detectors sensitivity thresholds 
and possible evolutionary effects, more luminous GRBs are those
more easily detectable at high redshifts (e.g., Amati 2006\nocite{Amati06}).
Besides the \epeiso{} correlation, which is fully satisfied by both GRB\,080916C and
GRB\,090323, the lack of accurate enough long term monitoring of the optical afterglow of these
events \citep{Greiner09,Kann09} prevents a stringent test of the correlations involving the break time
\tb{}.
However, we find that GRB\,080916C deviates by more than $\sim$2.5$\sigma$ from the best--fit
of the \epeisotb{} correlation (Fig.~5), suggesting that either the dispersion of 
this correlation is higher than thought before, or it is not satisfied at very high 
energies. This is an important issue, given that, with respect to the \epega{} 
correlation, the \epeisotb{} has the advantage, like the simple \epeiso{} correlation, 
of being model independent.

We expect that, thanks to \fermi{} and AGILE, the number of these extremely 
bright GRBs selected on the basis of their GeV emission 
will increase in the near future, 
giving us the possibility to better understand the physics of the prompt emission 
of GRBs and to get important clues on the reliability and origin of 
spectral energy correlations.

\begin{acknowledgements}

We thank Guido Barbiellini and Francesco Longo at $INFN$ (Trieste, Italy)
for discussions that prompted this work and Sara Cutini (at $ASI/ASDC$, 
Roma, Italy) for useful hints on the reduction of \fermi/GBM data.
\end{acknowledgements}


\begin{thebibliography}{}

\bibitem[Abdo et al. \ 2009]{Abdo09}
Abdo, A.A., Ackermann, M., Ajello, M., et al. 2009, Science, 323, 1688
%
\bibitem[Amati et al. \ 2002]{Amati02}
 Amati, L., Frontera, F., Tavani, M., et~al. 2002, A\&A, 390, 81
%
\bibitem[Amati \ 2006a]{Amati06}
 Amati, L. 2006a, MNRAS, 372, 233
%
\bibitem[Amati \ 2006b]{Amati06b}
 Amati, L. 2006b, N.Cim.B., 121, 1081
%
\bibitem[Amati et al. \ 2007]{Amati07}
 Amati, L., Della Valle, M., Frontera F., et~al. 2007, A\&A, 463, 913
%
\bibitem[Amati et al. \ 2008]{Amati08}
 Amati, L., Guidorzi, C., Frontera F., et~al. 2008, MNRAS, 391, 577
%
\bibitem[Amati \ 2008]{Amati08b}
 Amati, L. 2008, AIPC, 966, 3
%
\bibitem[Band et al. \ 1993]{Band93}
Band, D., Matteson, J., Ford, L. et al.,
ApJ, 413, 281
%
\bibitem[Band \& Preece \ 2005]{Band05}
Band, D., Preece, R.D. 2005,
ApJ, 627, 319
%
\bibitem[Barthelmy et al. \ 2008a]{Barthelmy08a}
Barthelmy, S.D., Baumgartner,W., Cummings, J., et al. 2008a, 
GCN Circ., 7606
%
\bibitem[Barthelmy et al. \ 2008b]{Barthelmy08b}
Barthelmy, S.D., Baumgartner,W., Cummings, J., et al. 2008b, 
GCN Circ., 8428
%
\bibitem[Baumgartner et al. \ 2008]{Baumgartner08}
Baumgartner, W., Barthelmy, S.D., Cummings, J., et al., 2008, 
GCN Circ., 8243
%
\bibitem[Bhat et al. \ 2008]{Bhat08}
Bhat, P.N., Preece, R.D., van der Horst, A.J. 2008, 
GCN Circ., 8526
%
\bibitem[Bissaldi et al. \ 2008a]{Bissaldi08a}
Bissaldi, E., McBreen, S., Wilson--Hodge, C.A., von Kienlin, A. 2008a, 
GCN Circ., 8263
%
\bibitem[Bissaldi et al. \ 2008b]{Bissaldi08b}
Bissaldi, E., McBreen, S., Connaughton, V. 2008b, 
GCN Circ., 8369
%
\bibitem[Bissaldi \& McBreen \ 2008]{Bissaldi08c}
Bissaldi, E., McBreen, S. 2008, 
GCN Circ., 8715
%
\bibitem[Butler et al. \ 2007]{Butler07}
Butler, N.R., Kocevski, D., Bloom, J.S., Curtis, J.L. 2007, 
ApJ, 671, 656
%
\bibitem[Butler et al. \ 2009]{Butler09}
Butler, N.R., Kocevski, D., Bloom, J.S. 2009, 
ApJ, 694, 76
%
\bibitem[Campana et al. \ 2007]{Campana07}
Campana, S., Guidorzi, C., Tagliaferri, G., et al. 2007, 
A\&A, 472, 395 
%
\bibitem[Chornock et al. \ 2009]{Chornock09}
Chornock, R., Perley, D.A., Cenko, S.B., Bloom, J.S. 2009, 
GCN Circ., 9028
%
\bibitem[Collazzi \& Schaefer \ 2008]{Collazzi08}
Collazzi, A.C., Schaefer, B.E. 2008,
ApJ, 688, 456
%
\bibitem[Connaughton \ 2009]{Connaughton09}
Connaughton, V. 2009, 
GCN Circ., 9230
%
\bibitem[Firmani et al. \ 2006]{Firmani06}
Firmani, C., Ghisellini, G., Avila--Rees, V., Ghirlanda, G. 2006, 
MNRAS, 370, 185 
%
\bibitem[Firmani et al. \ 2008]{Firmani08}
Firmani, C., Cabrera, J.I., Avila--Rees, V., et al. 2008, 
MNRAS, 393, 1209 
%
\bibitem[Frontera et al. in prep.]{Frontera09}
Frontera, F., et al. in prep.
%
\bibitem[Ghirlanda et al. \ 2004]{Ghirlanda04}
Ghirlanda, G., Ghisellini, G., Lazzati, D. 2004, 
ApJ, 616, 331 
%
\bibitem[Ghirlanda et al. \ 2005a]{Ghirlanda05a}
Ghirlanda, G., Ghisellini, G., Firmani, C. 2005a, 
MNRAS, 361, L10 
%
\bibitem[Ghirlanda et al. \ 2005b]{Ghirlanda05b}
Ghirlanda, G., Ghisellini, G., Firmani, C., Celotti, A., Bosnjak, Z. 2005b, 
MNRAS, 360, L45 
%
\bibitem[Ghirlanda et al.  \ 2006]{Ghirlanda06}
Ghirlanda, G., Ghisellini, G., Firmani, 2006, 
New J. Phys., 8, 123 
%
\bibitem[Ghirlanda et al. \ 2007]{Ghirlanda07}
Ghirlanda, G., Nava, L., Ghisellini, G., Firmani, C. 2007, 
A\&A, 466, 127 
%
\bibitem[Ghirlanda et al. \ 2008]{Ghirlanda08}
Ghirlanda, G., Nava, L., Ghisellini, G., Firmani, C., Cabrera J.I. 2008, 
MNRAS, 387, 319 
%
\bibitem[Ghirlanda et al. \ 2009]{Ghirlanda09}
Ghirlanda, G., Nava, L., Ghisellini, G., Celotti, A., Firmani C. 2009, 
A\&A, 496, 585 
%
\bibitem[Goldstein \& van der Horst \ 2008]{Goldstein08}
Goldstein, A. \& van der Horst, A. 2008, 
GCN Circ., 8245 
%
\bibitem[Golenetskii et al. \ 2007]{Golenetskii07}
Golenetskii, H., Aptekar, R., Mazets, E., et al. 2007, 
GCN Circ., 6849
%
\bibitem[Golenetskii et al. \ 2008a]{Golenetskii08a}
Golenetskii, H., Aptekar, R., Mazets, E., et al. 2008a, 
GCN Circ., 7751
%
\bibitem[Golenetskii et al. \ 2008b]{Golenetskii08b}
Golenetskii, H., Aptekar, R., Mazets, E., et al. 2008b, 
GCN Circ., 7812
%
\bibitem[Golenetskii et al. \ 2008c]{Golenetskii08c}
Golenetskii, H., Aptekar, R., Mazets, E., et al. 2008c, 
GCN Circ., 7854
%
\bibitem[Golenetskii et al. \ 2008d]{Golenetskii08d}
Golenetskii, H., Aptekar, R., Mazets, E., et al. 2008d, 
GCN Circ., 7862
%
\bibitem[Golenetskii et al. \ 2008e]{Golenetskii08e}
Golenetskii, H., Aptekar, R., Mazets, E., et al. 2008e, 
GCN Circ., 7995
%
\bibitem[Golenetskii et al. \ 2008f]{Golenetskii08f}
Golenetskii, H., Aptekar, R., Mazets, E., et al. 2008f, 
GCN Circ., 8258
%
\bibitem[Golenetskii et al. \ 2008g]{Golenetskii08g}
Golenetskii, H., Aptekar, R., Mazets, E., et al. 2008g, 
GCN Circ., 8548
%
\bibitem[Golenetskii et al. \ 2009a]{Golenetskii09a}
Golenetskii, H., Aptekar, R., Mazets, E., et al. 2009a, 
GCN Circ., 8776
%
\bibitem[Golenetskii et al. \ 2009b]{Golenetskii09b}
Golenetskii, H., Aptekar, R., Mazets, E., et al. 2009b, 
GCN Circ., 9030
%
\bibitem[Golenetskii et al. \ 2009c]{Golenetskii09c}
Golenetskii, H., Aptekar, R., Mazets, E., et al. 2009c, 
GCN Circ., 9050
%
\bibitem[Greiner et al. \ 2009]{Greiner09}
Greiner, J., Clemens. C., Kruehler, T., et al. 2009
A\&A, 498, 89 
%
\bibitem[Guiriec et al. \ 2009]{Guiriec09}
Guiriec, S., Connaughton, V., Briggs, M. 2009,
GCN Circ.,9336 
%
%\bibitem[Guidorzi et al. \ 2005]{Guidorzi05}
%Guidorzi, C., Frontera, F., Montanari, E., et al. 2005, 
%MNRAS, 363, 315 
%
%\bibitem[Guidorzi et al. \ 2006]{Guidorzi06}
%Guidorzi, C., Frontera, F., Montanari, E., et al. 2006, 
%MNRAS, 371, 843 
%
\bibitem[van der Horst \& Goldstein \ 2008]{Derhorst08}
van der Horst, A. \& Goldstein, A. 2008, 
GCN Circ., 8278
%
\bibitem[van der Horst \ 2009]{Derhorst09}
van der Horst, A. 2009, 
GCN Circ., 9035
%
\bibitem[Harrison et al. \ 2009]{Harrison09}
Harrison, F., Cenko, B., Frail, D.A.,
Chandra, P., Kulkarni, S. 2009, 
GCN Circ., 9043
%
\bibitem[Hurley et al. \ 2008]{Hurley08}
Hurley, K., Goldsten, J., Golenetskii, S., et al. 2008, 
GCN Circ.,8251
%
\bibitem[Kann et al. \ 2009]{Kann09}
Kann,D.A., Laux, U., Ludwig, F., Stecklum, S. 2009, 
GCN Circ., 9063
%
\bibitem[Kennea et al. \ 2009]{Kennea09}
Kennea, J., Evans, P., Goad, M. 2009, 
GCN Circ., 9024
%
\bibitem[von Kienlin \ 2009]{Kienlin09}
von Kienlin, A. 2009, 
GCN Circ., 9229
%
\bibitem[Lamb et al. \ 2004]{Lamb04}
Lamb, D. Q., Ricker, G. R., Atteia, J.-L., et al. 2004m
NewAR, 48, 423
%
\bibitem[Liang et al. \ 2004]{Liang04}
Liang, E., Dai, Z.G., Wu, X.F. 2004,
ApJ, 606, L29
%
\bibitem[Liang \& Zhang \ 2005]{Liang05}
Liang, E., Zhang, B. 2005,
ApJ, 633, 311
%
\bibitem[Li \ 2009]{Li09}
Li, Z. 2009,
ApJ, submitted [\rm arXiv:0810.2932]
%
\bibitem[Lloyd et al. \ 2000]{Lloyd00}
Lloyd, N.M., Petrosian, V., Mallozzi, R.S. 2000, 
ApJ, 534, 227 
%
\bibitem[Longo et al. \ 2009]{Longo09}
Longo, F., Moretti, E., Barbiellini, G., et al. 2009, 
GCN Circ., 9343
%
\bibitem[Meegan et al. \ 2008]{Meegan08}
Meegan, C.A., Greiner, J., Bhat, N.P., et al. 2008, 
GCN Circ., 8100
%
\bibitem[Nava et al. \ 2006]{Nava06}
Nava, L., Ghisellini, G., Ghirlanda, G., Tavecchio, F., Firmani, C. 2006
A\&A, 450, 471
%
\bibitem[Ohno et al. \ 2008]{Ohno08}
Ohno, M.,Kokubun, M.,Suzuki, M.,et al. 2008, 
GCN Circ., 7630
%
\bibitem[Ohno et al. \ 2009]{Ohno09a}
Ohno, M., Cutini, S., McEnery, J., et al. 2009, 
GCN Circ., 9021
%
\bibitem[Ohno \& Pelassa \ 2009]{Ohno09b}
Ohno, M., Pelassa, V. 2009, 
GCN Circ., 9334
%
\bibitem[Palmer et al. \ 2008a]{Palmer08a}
Palmer, D.M., Barthelmy, S.D., Baumgartner,W., et al. 2008a, 
GCN Circ., 8351
%
\bibitem[Palmer et al. \ 2008b]{Palmer08b}
Palmer, D.M., Barthelmy, S.D., Baumgartner,W., et al. 2008b, 
GCN Circ., 8526
%
\bibitem[Pal\'shin et al. \ 2008]{Palshin08}
Pal'shin, V., Golenetskii, S., Aptekar, R., et al. 2008, 
GCN Circ., 8256
%
\bibitem[Pal\'shin et al. \ 2009]{Palshin09}
Pal'shin, V., Golenetskii, S., Aptekar, R., et al. 2009, 
GCN Circ., 9196
%
%\bibitem[Reichart \ 2001]{Reichart01}
%Reichart, D.E. 2001, 
%ApJ, 553, 235 
%
\bibitem[Rau et al. \ 2009]{Rau09}
Rau, A., McBreen, S., Kruehler, T., Greiner, J. 2009, 
GCN Circ., 9353
%
\bibitem[Rossi et al. \ 2008]{Rossi08}
Rossi, F., Guidorzi, C., Amati, L., et al. 2008, 
MNRAS, 388, 1284 
%
\bibitem[Sakamoto et al. \ 2005]{Sakamoto05}
Sakamoto, T., Lamb, D.Q., Kawai, N., et al. 2005, 
ApJ, 629, 311
%
\bibitem[Sakamoto et al. \ 2008]{Sakamoto08}
Sakamoto, T., Barthelmy, S.D., Barbier, L., et al. 2008, 
ApJS, 175, 179
%
\bibitem[Shahmoradi \& Nemiroff \ 2009]{Shahmoradi09}
Shahmoradi, A., Nemiroff, R.J. 2009
MNRAS, submitted [\rm arXiv:0904.1464]
%
\bibitem[Stratta et al. \ 2008]{Stratta08}
Stratta, G., et al. 2008, 
GCN Report, 166.1
%
\bibitem[Tajima et al. \ 2008]{Tajima08}
Tajima, H., Bregeon, J., Chiang, J., et al. 2008, 
GCN Circ., 8246
%
\bibitem[Toma et al. \ 2005]{Toma05}
Toma, K., Yamazaki, R., Nakamura, T. 2004, 
ApJ, 635, 481
%
\bibitem[Updike et al. \ 2009]{Updike09}
Updike, A.C., Filgas, R.,
Kruehler, T., Greiner, J., McBreen, S. 2009, 
GCN Circ., 9026
%
\bibitem[Wang et al. \ 2009]{Wang09}
Wang, X.Y., Li, Z., Dai, Z.G., M\'esz\'aros, P. 2009,
ApJ, 698, L98
%
\bibitem[Yonetoku et al. \ 2004]{Yonetoku04}
Yonetoku, D., Murakami, T., Nakamura, T., et al. 2004, 
ApJ, 609, 935 
%
\bibitem[Zhang \& M\'esz\'aros \ 2002]{Zhang02}
Zhang, B., M\'esz\'aros, P. 2002,
ApJ, 581, 1236
%
\end{thebibliography}
\end{document}